\documentclass{svjour3}                     
\smartqed  
\usepackage{graphicx}

\usepackage{latexsym}

\journalname{Hyperfine Interactions}

\begin{document}

\title{Radiative Corrections and $Z^\prime$}

\author{Jens Erler}

\institute{Jens Erler \at
              Departamento de F\'isica Te\'orica, Instituto de F\'isica, \\
              Universidad Nacional Aut\'onoma de M\'exico, 04510 M\'exico D.F., M\'exico \\
              Tel.: +52-55-5622-5166 \\
              Fax: +52-55-5622-5015 \\
              \email{erler@fisica.unam.mx} \\
              \emph{Present address:} Institut f\"ur Theoretische Physik E, RWTH Aachen,
              52056 Aachen, Germany}

\date{Received: date / Accepted: date}

\maketitle

\begin{abstract}
Radiative corrections to parity violating deep inelastic electron scattering (PVDIS) are reviewed including a discussion of the renormalization group evolution (RGE) of the weak mixing angle. Recently obtained results on hypothetical 
$Z^\prime$ bosons --- for which parity violating observables play an important r\^ole --- are also presented.
\end{abstract}

\section{Effective electroweak interactions}
\label{intro}
The first two terms of the Lagrangian, 
${\cal L} = {\cal L}_{\rm fermion} + {\cal L}_{\rm Yukawa} + {\cal L}_{\rm gauge} + {\cal L}_{\rm Higgs}$, 
of the electroweak Standard Model (SM) contain the free fermionic part and the interactions,
$${\cal L}_A + {\cal L}_{\rm W} + {\cal L}_{\rm Z} = - {g\over 2} \left(2 \sin^2\theta_W J^\mu_A A_\mu + 
J^\mu_W W^-_\mu + J^{\mu\dagger}_W W^+_\mu + {1\over \cos^2\theta_W} J^\mu_Z Z_\mu \right),$$
in terms of the electromagnetic current, 
$J^\mu_A = \sum\limits_{i = 1}^3 
\left( {2\over 3} \bar{u}^i \gamma^\mu u^i - {1\over 3} \bar{d}^i \gamma^\mu d^i - \bar{e}^i \gamma^\mu e^i \right)$, \\ 
the weak charged current (CC), 
$J^{\mu \dagger}_W = \sqrt{2} \sum\limits_{i = 1}^3 
\left( \bar{u}^{i0} \gamma^\mu P_L d^{i0} + \bar{\nu}^{i0} \gamma^\mu P_L e^{i0} \right)$, \\ 
and the weak neutral current (NC), 
$J^\mu_Z \equiv \sum\limits_{i=1}^{N_\psi} \bar\psi^i \gamma^\mu [g_V^i - g_A^i \gamma^5] \psi^i = 
- 2 \sin^2\theta_W J^\mu_A + \sum\limits_{i = 1}^3 \left( \bar{u}^i \gamma^\mu P_L u^i - \bar{d}^i \gamma^\mu P_L d^i + 
\bar{\nu}^i \gamma^\mu P_L\nu^i - \bar{e}^i \gamma^\mu P_L e^i \right)$, 
where $P_L \equiv {1 - \gamma^5\over 2}$. 
At the tree-level, the NC couplings,
$g_V^i = {1\over 2} \tau_3^{ii} - 2 Q^i \sin^2\theta_W$ and $g_A^i = {1\over 2} \tau_3^{ii}$,
with $Q^i$ ($\tau_3$) denoting the electric charge (third Pauli matrix), give rise to the effective 4-Fermi Hamiltonian,
$${\cal H}_{\rm eff} = {1\over 2} \left( {g\over 2 \cos^2\theta_W M_Z} \right)^2 J^\mu_Z J_{\mu Z} = 
{G_F\over \sqrt{2}} J^\mu_Z J_{\mu Z} = 
{G_F\over \sqrt{2}} \sum\limits_{MNij} h_{MN}^{ij} \bar{\psi}^i \Gamma^M \psi^i \bar{\psi}^j \Gamma^N \psi^j, $$
where $\Gamma^V = \gamma^\mu$, $\Gamma^A = \gamma^\mu \gamma^5$, and $h_{MN}^{ij} = g_M^i g_N^j$. 
Unfortunately, there is no generally accepted notation, normalization, and sign convention for the $h_{MN}^{ij}$ in the literature. For parity violating $eq$ interactions one defines 
$C_{1q} \equiv 2 h_{AV}^{eq}$ and $C_{2q} \equiv 2 h_{VA}^{eq}$. 
Parity violation in heavy atoms~\cite{Budker} is basically driven by the $C_{1q}$, while PVDIS~\cite{Souder} determines approximately the combination,
$\omega_{\rm PVDIS} \equiv(2\ C_{1u} - C_{1d})+0.84\ (2\ C_{2u} - C_{2d})$.

\section{Radiative corrections}
Including one-loop electroweak radiative corrections one obtains the expressions~\cite{Marciano:1982mm},
$$
2\ C_{1u} - C_{1d} = - {3\over 2} \left[ \rho_{\rm NC} - {\alpha\over 2\pi} \right] \left[ 1 - {20\over 9} 
\left( \sin^2\hat\theta_W(0) -{ 2\alpha\over 9\pi} \right) \right] + \Box_{WW} + \Box_{ZZ} + \Box_{\gamma Z}
$$
\begin{equation} 
\label{eq:c1}
+ {5\hat\alpha\over9\pi} [1 - 4 \sin^2\hat\theta_W(M_Z)] \left[ \ln {M_Z\over m_e} + {1\over 12} \right], 
\end{equation}
$$
2\ C_{2u} - C_{2d} = - {3\over 2} \left[ \rho_{\rm NC} - {\alpha\over 6\pi} \right] \left[ 1 - 4 
\left( \sin^2\hat\theta_W(0) - {2\alpha\over 9\pi} \right) \right] +\Box_{WW} + \Box_{ZZ} + \Box_{\gamma Z}
$$
\begin{equation} 
\label{eq:c2}
+ {5\hat\alpha\over 9\pi} [1 - {12\over 5} \sin^2\hat\theta_W(M_Z)] \left[ \ln {M_Z\over m_q} + {1\over 12} \right]
- {8\hat\alpha\over 9\pi} \left[ \ln {M_W\over m_q} + {1\over 12} \right], 
\end{equation}
where $ \rho_{\rm NC} \approx 1.0007$ collects various propagator and vertex corrections relative to $\mu$-decay, and the second lines are from the $e$ and $q$ charge radii. With $\hat{s}^2 \equiv \sin^2\hat\theta_W(M_Z)$,
$$ 
\Box_{WW} = - {9 \hat\alpha\over 8\pi \hat{s}^2} \left[ 1 - {\hat\alpha_s(M_W)\over 3\pi} \right], 
\Box_{\gamma Z} = - {3 \hat\alpha\over 4\pi} [ 1 - 4\hat{s}^2] \left[ \ln {M_Z\over M_\rho} + {3\over 4} \right] ,
\Box_{ZZ}  \ll \Box_{WW} 
$$
are the box contributions except that for $2\ C_{2u} - C_{2d}$ the $\alpha_s$ correction to the $WW$-box~\cite{Erler:2003yk},  $\Box_{WW}$, is not yet known and $\Box_{\gamma Z}$ is obtained from above by replacing $4 \hat{s}^2$ by $28 \hat{s}^2/9$ and the constant 3/4 by 5/12. The numerical results are summarized in Table~\ref{tab:PVDIS}.

\begin{table}
\caption{Numerical contributions to $\omega_{\rm PVDIS}$}
\label{tab:PVDIS}    
\begin{tabular}{llll}
\hline\noalign{\smallskip}
& $2\ C_{1u} - C_{1d}$ &$2\ C_{2u} - C_{2d}$ & $\omega_{\rm PVDIS}$ \\
\noalign{\smallskip}\hline\noalign{\smallskip}
tree + QED & $-0.7060$ & $-0.0715$ & $-0.7660$ \\
charge radii & +0.0013 & $-0.0110$ & $-0.0079$ \\
$\Box_{WW}$ & $-0.0120$ & $-0.0120$ & $-0.0220$ \\
$\Box_{\gamma Z}$ & $-0.0008$ & $-0.0027$ & $-0.0031$ \\
other & $-0.0009$ & $-0.0011$ & $-0.0018$ \\
\noalign{\smallskip}\hline\noalign{\smallskip}
TOTAL & $-0.7184$  & $-0.0983$ & $-0.8010$ \\
\noalign{\smallskip}\hline
\end{tabular}
\end{table}

Eqs.~(\ref{eq:c1}) and (\ref{eq:c2}) were originally obtained for atomic parity violation.  For PVDIS, the one-loop expressions with the full kinematic dependance (in analogy with Ref.~\cite{Czarnecki:1995fw} for polarized M\o ller scattering) need to be computed, plus the $\hat\alpha_s$ corrections to $\Box_{WW}$ and $\Box_{ZZ}$. In practice, one would want to define new $C_{2q}$ at these kinematics since these would supersede the ones at very low $Q^2$ with their large hadronic uncertainties.

\begin{figure}
\includegraphics[width=1.0\textwidth]{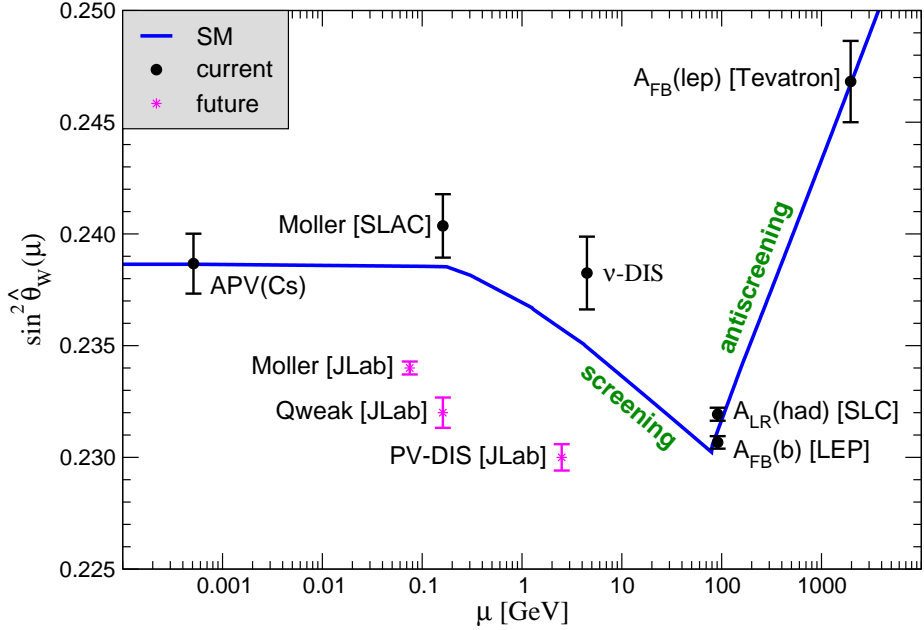}
\caption{Renormalization scale dependance of the weak mixing angle. Various measurements are shown at their nominal energy scales, {\em i.e.}, not necessarily at their typical momentum transfers.}
\label{fig:running_s2w} 
\end{figure}

The $\overline{\rm MS}$ scheme (marked by a caret) weak mixing angle enters Eqs.~(\ref{eq:c1}) and (\ref{eq:c2}) evaluated at the renormalization scale $\mu = 0$. Introducing the quantity 
$\hat{X} \equiv \sum_i N_C^i \gamma^i \hat g_V^i Q^i$, where $N_C = 3$ (1) for quarks (leptons) and $\gamma^i = 4$ (22) for chiral fermions (gauge bosons), one can show that $d\hat{X}/X = d\hat{\alpha}/\alpha$, {\em i.e.}, the RGE for $\hat\alpha(\mu)$ implies that for $\sin^2\hat\theta_W(\mu)$ (see Fig.~\ref{fig:running_s2w}) including experimental constraints from $e^+ e^-$ annihilation and $\tau$ decays that enter the dispersion integral for the non-perturbative regime, {\em provided\/} that any one of the following conditions is satisfied: (i) no mass threshold is crossed; 
(ii) perturbation theory applies ($W^\pm$, leptons, $b$ and $c$ quarks); (iii) equal coefficients (like for $d$ {\em vs.\/} $s$ quarks); or (iv) symmetries like $SU(2)_I$ or $SU(3)_F$ may be applied. 

This leaves as the only problem area the treatment of the $u$ {\em vs.\/} the ($d$, $s$) quark thresholds, or ---considering that $m_s \neq m_d \approx m_u$ --- the separation of the $s$ quarks from the ($u$, $d$) doublet. Our strategy~\cite{Erler:2004in} is to define threshold masses (absorbing QCD matching effects), 
$\bar{m}_q = \xi_q M_{1S}/2$, in terms of $1S$ resonance masses. The $\xi_q$ are between 0 (chiral limit) and 1 (infinitely heavy quarks). One expects $\xi_b >\xi_c >\xi_s >\xi_d >\xi_u$ and we explicitly verified $\xi_b >\xi_c$ in perturbative QCD. Now, $\xi_s = \xi_c$ defines the heavy quark limit for the $s$ quark, implying $\bar{m}_s < 387$~MeV.
On the other hand, $\xi_s = \xi_d \approx \xi_u$ together with the dispersion result for the three-flavor RGE for 
$\hat\alpha$ below $\mu = \bar{m}_c$, $\Delta\hat\alpha^{(3)} (\bar{m}_c)$, yields an upper limit on the $s$ quark contribution and $\bar{m}_s > 240$~MeV. Besides parametric uncertainties from the input values of $\hat{m}_b$, 
$\bar{m}_c$, and $\hat\alpha_s$, this procedure introduces an experimental error through 
$\Delta\hat\alpha^{(3)} (\bar{m}_c)$ ($\pm 3\times 10^{-5}$), $SU(3)_F$ breaking masses, 
$\bar{m}_u = \bar{m}_d \neq \bar{m}_s$ ($\pm 5\times 10^{-5}$), and $SU(2)_I$ breaking masses, 
$\bar{m}_u \neq \bar{m}_d$ ($\pm 8\times 10^{-6}$). Starting at three-loop order there is also the (OZI rule violating) singlet (QCD annihilation) contribution to the RGE for $\hat\alpha$ (but by virtue of 
$Q_u + Q_d + Q_s = \tau^{uu}_3 + \tau^{dd}_3 = 0$ not present in $\hat{X}$) introducing another $\pm 3\times 10^{-5}$ error.

\section{$Z^\prime$ physics: the search for a fifth force}
Extra $Z^\prime$ bosons are predicted in virtually all scenarios for TeV scale physics beyond the SM, including grand unified theories, left-right models, superstrings, technicolor, large extra dimensions and little Higgs theories and in all these cases one expects $M_{Z^\prime} =  {\cal O}({\rm TeV})$ and 100 (1,000)~fb$^{-1}$ of LHC data will explore $M_{Z^\prime}$ values up to 5 (6)~TeV~\cite{Godfrey:2002tna}. Angular distributions of leptons may help to discriminate spin-1 ($Z^\prime$) against spin-0 (sneutrino) and spin-2 (Kaluza-Klein graviton) resonances~\cite{Osland:2009tn}. The LHC will also have some diagnostic tools to narrow down the underlying $Z'$ model by studying, {\em e.g.,} leptonic forward-backward asymmetries and heavy quark final states~\cite{Barger:2006hm,Godfrey:2008vf}.

$Z^\prime$ models based on the gauge group $E_6$ without kinetic mixing correspond to extending the SM by a $U(1)^\prime = \cos\beta\ U(1)_\chi + \sin\beta\ U(1)_\psi$ ($-90^\circ < \beta \leq 90^\circ$). 
Particular values for $\beta$ give $Z^\prime$ models of special interest, namely 
(i) $\beta = 0^\circ \Longrightarrow Z_\chi$ and is defined by the breaking of $SO(10) \to SU(5) \times U(1)_\chi$; 
(ii) $\beta = 90^\circ \Longrightarrow Z_\psi$ defined by the breaking of $E_6 \to SO(10) \times U(1)_\psi$; 
(iii) $\beta \approx - 52.2^\circ \Longrightarrow Z_\eta$ and appears in a class of heterotic string models compactified on Calabi-Yau manifolds; 
(iv) $\beta \approx 37.8^\circ \Longrightarrow Z_I \perp Z_\eta$ and is hadrophobic in that it doesn't couple to up-type quarks; 
(v) $\beta \approx 23.3^\circ \Longrightarrow Z_S$ and gives rise to the so-called secluded $U(1)^\prime$ breaking model addressing both the little hierarchy problem ($M_Z \ll M_{Z^\prime}$)~\cite{Erler:2002pr} and electroweak baryogenesis~\cite{Kang:2004pp}; and 
(vi) $\beta \approx 75.5^\circ \Longrightarrow Z_N$ with no couplings to right-handed neutrinos and therefore allowing the (ordinary) see-saw mechanism. Adding kinetic mixing is equivalent to considering the more general combination, 
$Z^\prime = \cos\alpha \cos\beta\ Z_\chi + \sin\alpha \cos\beta\ Z_Y + \sin\beta\ Z_\psi.$ 
Then the values 
(vii) $(\alpha,\beta) \approx (50.8^\circ, 0^\circ) \Longrightarrow Z_R$ defined by the breaking of $SU(2)_R \to U(1)_R$; 
(viii) in left-right symmetric models appears the $Z_{LR} \propto 1.53\ Z_R - 0.33\ Z_{B-L}$, where
 $(\alpha,\beta) \approx (-39.2^\circ, 0^\circ) \Longrightarrow Z_{B-L} \perp Z_R$; while
(ix) $(\alpha,\beta) \approx (28.6^\circ, -48.6^\circ) \Longrightarrow Z_{\not{L}}$ with no couplings to charged leptons and left-handed neutrinos. Finally, 
(x) the sequential $Z_{SM}$ couples like and could be an excited state of the ordinary $Z$ boson.

$Z^\prime$ bosons can have various effects on precision observables. The $Z$-$Z^\prime$ mixing angle, 
$\theta_{ZZ^\prime}$, is strongly constrained by the $M_W$-$M_Z$ interdependence (even for the $Z_{\not{L}}$) and
by the $Z$-pole (because $\theta_{ZZ^\prime}$ affects the very precisely measured $Z$ couplings to fermions). Conversely, if $\theta_{ZZ^\prime} = 0$ the $Z$ pole observables are rather blind to $Z^\prime$ physics because the $Z$ and $Z^\prime$ amplitudes are almost completely out of phase and one needs to go off-peak, {\em i.e.}, to LEP~2 and low energies. There are also loop effects which are small but not necessarily negligible. {\em E.g.}, the $M_W$-$G_F$ relation, parametrized by $\Delta \hat{r}_W$, is shifted,
\begin{equation}
\label{deltar}
\delta (\Delta \hat{r}_W) = - {5\over 2} {\alpha\over \pi \cos^2\theta_W} \lambda \epsilon_L^e \epsilon_L^\mu 
{M_W^2\over M_{Z^\prime}^2 - M_W^2} \ln {M_{Z^\prime}^2\over M_W^2},
\end{equation}
where the $\epsilon_L^f$ denote $U(1)^\prime$ charges and $\lambda$ is a model dependent parameter of ${\cal O}(1)$.
$Z^\prime$ bosons would also yield an apparent violation of first row CKM unitarity, 
$\delta( V_{ud}^2 + V_{us}^2 + V_{ub}^2)$, given by the r.h.s.\ of Eq.~(\ref{deltar}) upon replacing $\epsilon_L^\mu$ by $-2 (\epsilon_L^\mu - \epsilon_L^d)$. Finally, the muon anomalous magnetic moment~\cite{Hertzog} would receive a (usually tiny) correction, $\delta a_\mu = 5 /36\ \alpha /\pi \cos^2\theta_W\ \lambda (V_\mu^2 - 5 A_\mu^2)\ m_\mu^2/M_{Z^\prime}^2$, with some interest for the $Z_\psi$ which is insensitive to most other precision data (since it does not possess any vector couplings $V_f$) while the axial coupling $A_\mu$ comes enhanced in $\delta a_\mu$.

Results from a global analysis~\cite{Erler:2009jh} are shown in Table~\ref{tab:limits}. Some $Z'$ models give a fairly low minimum $\chi^2$, especially the $Z_\psi$ and $Z_R$. Technically, there is a 90\% C.L.\ {\em upper\/} bound on the $Z_R$ mass of about 29 TeV. Of course, at present there is little significance to this observation since there are two additional fit parameters ($M_Z'$ and $\theta_{ZZ'}$) and various adjustable charges (like the angles $\alpha$ and $\beta$). Still this surprises given that the SM fit is quite good with $\chi^2_{\rm min} = 48.0/45$ (with $M_H$ unconstrained). It is interesting that the improvement, $\Delta\chi^2_{\rm min} = - 2.9$, is mainly from PAVI observables, namely from polarized M\o ller~\cite{Anthony:2005pm} ($-1.7$) and $e^-$-hadron scattering~\cite{Young:2007zs} ($-0.9$). The best fit with $M_{Z'} = 667$~GeV implies shifts in the so-called weak charges, $\delta |Q_W(e,p)| = - 0.0073$, corresponding to $6.6\sigma$ and $2.5\sigma$, respectively, for the proposed MOLLER~\cite{Kumar} and Qweak~\cite{Page} experiments at JLab. Similarly, expect $\delta |\omega_{\rm PVDIS}|= - 0.0200$ ($4.2\sigma$). 
 
\begin{table}
\caption{95\% C.L.\ lower mass limits (in GeV) on extra $Z'$ bosons and lower and upper limits for $\theta_{ZZ'}$ from electroweak precision data, assuming 114.4~GeV~$ < M_H < 1$~TeV. Also shown are for comparison (where applicable) the limits obtained by CDF (they assume that no supersymmetric or exotic decay channels are open; otherwise the limits would be moderately weaker) and LEP~2 (constraining virtual $Z'$ bosons by their effects on cross sections and angular distributions of di-leptons, hadrons, $b\bar{b}$ and $c\bar{c}$ final states). CDF sees a significant excess at a di-electron invariant mass of 240~GeV, but this is not confirmed in the $\mu^+ \mu^-$ channel. The result for the leptophobic $Z_{\not{L}}$ (in parentheses) in the electroweak column assumes a specifically chosen Higgs sector.  The CDF number refers to the $Z_{SM}$ limit from the di-jet channel and should give a rough estimate of the sensitivity to our specific $Z_{\not{L}}$. The various mass limits are highly complementary ({\em e.g.,} unlike Tevatron limits, electroweak and LEP~2 limits scale with the coupling strength). The last column indicates the $\chi^2$ minimum for each model.}
\label{tab:limits}
\begin{tabular}{lrrrrrccc} 
\hline\noalign{\smallskip}
 $Z'$ & electroweak & CDF & LEP 2 & $\theta_{ZZ'}^{\rm min}$ & $\theta_{ZZ'}^{\rm max}$ & $\chi^2_{\rm min}$ \\
\noalign{\smallskip}\hline\noalign{\smallskip}
$Z_\chi$         & 1,141\phantom{OOO} &    892 &    673 & $-0.0016$ & 0.0006 & 47.3 \\ 
$Z_\psi$         &    147\phantom{OOO} &    878 &    481 & $-0.0018$ & 0.0009 & 46.5 \\ 
$Z_\eta$         &    427\phantom{OOO} &    982 &    434 & $-0.0047$ & 0.0021 & 47.7 \\ 
$Z_I$              & 1,204\phantom{OOO} &    789 &           & $-0.0005$ & 0.0012 & 47.4 \\ 
$Z_S$            & 1,257\phantom{OOO} &    821 &            & $-0.0013$ & 0.0005 & 47.3 \\ 
$Z_N$            &    623\phantom{OOO} &    861 &            & $-0.0015$ & 0.0007 & 47.4 \\
$Z_R$            &    442\phantom{OOO} &           &            & $-0.0015$ & 0.0009 & 46.1 \\ 
$Z_{LR}$       &    998\phantom{OOO} &    630 &      804 & $-0.0013$ & 0.0006 & 47.3 \\
$Z_{\not{L}}$ &   (803)\phantom{IOO} & (740) \hspace{-6pt} & & $-0.0094$ & 0.0081 & 47.7 \\ 
$Z_{SM}$      & 1,403\phantom{OOO} & 1,030 & 1,787 & $-0.0026$ & 0.0006 & 47.2 \\
\noalign{\smallskip}\hline
\end{tabular}
\end{table}

\begin{acknowledgements}
It is a pleasure to thank Paul Langacker, Shoaib Munir and Eduardo Rojas for collaboration. This work is supported by CONACyT project 82291--F. 
\end{acknowledgements}

\end{document}